\pdfoutput=1
\documentclass[preprint,aps]{revtex4-1}
\usepackage{graphicx}
\usepackage{rotating}
\usepackage{float}
\newcommand{\be}{\begin{equation}}
\newcommand{\ee}{\end{equation}}
\newcommand{\bea}{\begin{eqnarray}}
\newcommand{\eea}{\end{eqnarray}}
\newcommand{\ba}{\begin{array}}
\newcommand{\ea}{\end{array}}

\begin{document}
\title{Axial-vector form factors for the low lying octet baryons in the chiral quark constituent model}
\author{Harleen Dahiya}
\affiliation{Department of Physics,\\ Dr. B.R. Ambedkar National
Institute of Technology,\\ Jalandhar, 144011, India}
 \author{Monika Randhawa}
 \affiliation{University Institute of Engineering and Technology,
Panjab University, Chandigarh, India}
\begin{abstract}
We have calculated the axial-vector form factors of the low lying octet baryons ($N$, $\Sigma$, $\Xi$ and $\Lambda$) in
the chiral constituent quark model ($\chi$CQM).
In particular, we have studied the implications of chiral symmetry breaking and SU(3) symmetry
breaking for the singlet ($g^0_{A}$) and non-singlet ($g^3_{A}$ and $g^8_{A}$) axial-vector coupling constants expressed as combinations of the spin polarizations at zero momentum transfer. The conventional dipole form of parametrization has been used to analyse the $Q^2$ dependence of the axial-vector form factors ($G^0_{A}(Q^2)$, $G^3_{A}(Q^2)$ and $G^8_{A}(Q^2)$). The total strange singlet and non-singlet
contents ($G_s^0(Q^2)$, $G_s^3(Q^2)$ and $G_s^8(Q^2)$) of the nucleon determining the strange quark contribution to the nucleon spin ($\Delta s$) have also been discussed.

\end{abstract}

\maketitle
\section{Introduction}
The internal structure of the baryons has been extensively studied
ever since the measurements of polarized structure functions of proton in the deep
inelastic scattering (DIS) experiments \cite{emc,smc,adams,hermes}. These experiments have provided the first evidence that the valence quarks of proton carry only a small fraction of its spin and the decomposition of the proton's spin still remains to be a major unresolved issue in high energy spin physics. Form factors parameterized from the electromagnetic current operator as well as the isovector axial-vector current operator are important in hadron physics as they provide a deep insight in understanding the internal structure. The electromagnetic Dirac and Pauli form factors are well know over a wide region of momentum transfer squared $Q^2$, however, the study of the axial-vector form factors has been rather limited. The measured first moment is related to the combinations of the axial-vector coupling constants which are combinations of the spin polarizations, $\Delta u$, $\Delta d$ and $\Delta s$. For example,
\be
\Gamma_1^p(Q^2)=\int_0^1 g_1^p(x,Q^2) dx=\frac{C_s(Q^2)}{9} g_A^0+\frac{C_{ns}(Q^2)}{12} g_A^3+\frac{C_{ns}(Q^2)}{36} g_A^8.
\ee
Here $C_s$ and $C_{ns}$  are the flavor singlet and non-singlet Wilson coefficients calculable from perturbative QCD. The quantity $g_A^0$ corresponds to the flavor singlet component related to the total quark spin content $\Delta \Sigma$ whereas $g_A^3$ and  $g_A^8$ correspond to the flavor
non-singlet components usually obtained from the neutron $\beta-$decay and the semi-leptonic weak decays
of hyperons respectively. These axial-vector coupling constants can be related to certain well known sum rules such as
Bjorken sum rule (BSR) \cite{bjorken} and
Ellis-Jaffe sum rule (EJSR) \cite{ellis} and derived within Quantum
Chromodynamics (QCD) using operator product expansion, renormalization group invariance and
isospin conservation in the DIS.

Recently, experiments measuring electromagnetic and weak form factors from the elastic scattering of electrons, for
example, SAMPLE at MIT-Bates \cite{sample}, G0 at JLab \cite{g0},
PVA4 at MAMI \cite{a4} and HAPPEX  at JLab \cite{happex} have given indications of strangeness contribution in the nucleon. These
experiments have provided considerable insight on the role played
by strange quarks in the charge, current and spin structure of the nucleon.
The nucleon axial coupling constant $g_A^3$ has received much attention in the past and has been determined precisely from nuclear $\beta-$decay \cite{PDG}. It corresponds to the value of the axial form factor at zero-momentum transfer ($Q^2 = -q^2 = 0$). It is one of  the fundamental parameter to test the chiral symmetry breaking effects and thereby determine the spin structure of the
nucleon.  Our information about the other low lying octet baryon axial-vector form factors from
experiment is also rather limited because it is difficult to
measure the hyperon properties experimentally due to their
short lifetimes. Even though there has been considerable
progress in the past few years to determine the $Q^2$ dependence of axial form factors experimentally, there is no consensus regarding the various mechanisms which can contribute to it. Experiments involving elastic  scattering of neutrinos and antineutrinos \cite{antineutrino1,antineutrino2} and the pion electro-production on the proton \cite{pion-electro} have explored $Q^2$ dependence of axial form factors in the past and they point out the need for additional refined data. More recently, there has been considerable refinement to measure the $Q^2$ dependence of the axial-vector form factor of the nucleon in the higher-energy Miner$\nu$a experiment
at Fermilab \cite{minerva}.

The theoretical knowledge in this regard has been rather limited because of confinement and it is still a big challenge
to perform the calculations from the first principles of QCD. Even though some lattice QCD calculations of
the axial charge and form factors of the nucleon have
been performed \cite{lattice} but still a lot of refinements need to be done. The broader question of axial charge, axial form factors and the strange quark contribution to the axial form factors of the nucleon has also been discussed by several
authors in other models recently \cite{authors}.
In addition to this, the partial conservation of axial-vector current (PCAC) also provide important constraints on the axial exchange currents  to describe the non-valence degrees of freedom in the nucleon \cite{pcac-ref1,pcac-ref2,buchmann-axial-mass}.
One of the most successful nonperturbative approach which finds its
application for the quantities discussed above is the chiral constituent
quark model ($\chi$CQM) \cite{manohar}.  The basic idea is based on
the possibility that chiral symmetry breaking takes place at
a distance scale much smaller than the confinement scale.
The $\chi$CQM uses the effective interaction Lagrangian
approach of the strong interactions  where the effective degrees of freedom are the
valence quarks and the internal Goldstone bosons (GBs)
which are coupled to the valence quarks \cite{cheng,johan,song,hd}. The $\chi$CQM successfully explains  the ``proton
spin problem'' \cite{hd}, magnetic moments of
octet and decuplet baryons including their
transitions \cite{hdmagnetic}, account for the violation of Gottfried Sum Rule
\cite{hdasymmetry} and Coleman-Glashow sum rule, hyperon $\beta$ decay parameters
\cite{nsweak}, strangeness content in the nucleon \cite{hds},
magnetic moments of ${\frac{1}{2}}^-$ octet
baryon resonances \cite{nres}, magnetic moments of ${\frac{1}{2}}^-$ and ${\frac{3}{2}}^-$
$\Lambda$ resonances \cite{torres}, charge radii \cite{charge-radii}, quadrupole moment \cite{quad}, etc..
The model is successfully extended to
predict the important role played by the small intrinsic charm content in the nucleon spin in the SU(4) $\chi$CQM \cite{hdcharm} and to
calculate the magnetic moment and charge radii of spin
${\frac{1}{2}}^+$ and spin ${\frac{3}{2}}^+$ charm baryons including
their radiative decays \cite{nscharm,chargeradii-charm}. The $\chi$CQM provides a simultaneous unique information on the flavor and spin structure of the baryons including the heavy baryons.
In view of the above
developments in the $\chi$CQM, it become desirable to extend the
model to calculate the axial-vector form factors of the low-lying octet baryons. It is widely recognized that a knowledge
about the axial-vector form factors of the baryons in general and the strangeness content of the nucleon in particular would
undoubtedly provide vital clues to the nonperturbative
aspects of QCD.

The purpose of the present communication is to determine the
axial-vector form factors of the low lying octet baryons in
the chiral constituent quark model ($\chi$CQM).
In particular, we
would like to phenomenologically estimate the quantities affected
by chiral symmetry breaking and SU(3) symmetry
breaking.
We begin by computing the static properties of the axial-vector current.
The singlet ($g^0_{A}$) and non-singlet ($g^3_{A}$ and $g^8_{A}$) axial-vector coupling constants expressed as combinations of the spin polarizations at zero momentum transfer have been investigated for the case of $N$, $\Sigma$, $\Xi$ and $\Lambda$ baryons.
Further, it would be significant to analyse the $Q^2$ dependence of the axial-vector form factors ($G^0_{A}(Q^2)$, $G^3_{A}(Q^2)$ and $G^8_{A}(Q^2)$) as well as their explicit flavor contributions ($G^u_{A}(Q^2)$, $G^d_{A}(Q^2)$ and $G^s_{A}(Q^2)$) by using a conventional dipole form of parametrization.
Furthermore, it would be interesting to extend the calculations to predict the
total strange singlet and non-singlet
contents ($G_s^0(Q^2)$, $G_s^3(Q^2)$ and $G_s^8(Q^2)$) of the nucleon and determine the strange quark contribution to the nucleon spin ($\Delta s$). The results can be compared with the recent available experimental observations.

\section{Chiral Constituent Quark Model}
The key to understand the
structure of the baryons, in the $\chi$CQM formalism \cite{cheng},
is the fluctuation process
\be q^{\pm} \rightarrow {\rm GB} + q^{'
\mp} \rightarrow (q \bar q^{'}) +q^{'\mp}\,, \label{basic}
\ee
where GB represents the Goldstone boson and $q \bar q^{'} +q^{'}$
constitute the ``quark sea'' \cite{cheng,johan,hd}. The effective
Lagrangian describing the interaction between quarks and a nonet
of GBs,  can be expressed as
\be
{\cal L}= c_8 {\bf \bar
q}\left(\Phi+\zeta\frac{\eta'}{\sqrt 3}I \right) {\bf q}=c_8 {\bf
\bar q}\left(\Phi' \right) {\bf q}\,,\label{lag}
\ee where $\zeta=c_1/c_8$, $c_1$ and $c_8$ are the coupling constants for
the singlet and octet GBs, respectively, $I$ is the $3\times 3$
identity matrix. The matrix ${\bf {\rm}q}$ and the GB field can be expressed in terms of the GBs and their transition probabilities as \bea {\bf {\rm}q} =\left( \ba{c} u \\ d \\ s \ea \right),~~~~
\Phi' = \left( \ba{ccc} \frac{\pi^o}{\sqrt 2}
+\beta\frac{\eta}{\sqrt 6}+\zeta\frac{\eta^{'}}{\sqrt 3} & \pi^+
  & \alpha K^+   \\
\pi^- & -\frac{\pi^o}{\sqrt 2} +\beta \frac{\eta}{\sqrt 6}
+\zeta\frac{\eta^{'}}{\sqrt 3}  &  \alpha K^o  \\
 \alpha K^-  &  \alpha \bar{K}^o  &  -\beta \frac{2\eta}{\sqrt 6}
 +\zeta\frac{\eta^{'}}{\sqrt 3} \ea \right). \eea
If the parameter
$a(=|c_8|^2$) denotes the transition probability of chiral
fluctuation of the splitting $u(d) \rightarrow d(u) + \pi^{+(-)}$,
then $\alpha^2 a$, $\beta^2 a$ and $\zeta^2 a$ respectively,
denote the probabilities of transitions of $u(d) \rightarrow s + K^{+(0)}$, $u(d,s)
\rightarrow u(d,s) + \eta$, and $u(d,s) \rightarrow u(d,s) +
\eta^{'}$ \cite{cheng,johan}. These parameters provide the basis to
understand the extent to which the quark sea contributes to the
structure of the baryon. The symmetry breaking parameter $a$ is
introduced by considering nondegenerate quark masses $M_s > M_{u,d}$, the parameters $\alpha$ and $\beta$ are introduced by considering nondegenerate GB masses $M_{K,\eta}> M_{\pi}$ and finally the parameter $\zeta$ is introduced by considering  $M_{\eta^{'}} > M_{K,\eta}$. Since the quark
contributions scale as $\frac{1}{M_q^2}$, a  hierarchy for the probabilities can be obtained as
\be a> a \alpha^2 \geq a \beta^2> a \zeta^2.
\ee

Before proceeding further to calculate the axial-vector form factors, we briefly discuss the calculation of
the spin structure of the baryons. Following references {\cite{{cheng},{johan}}},
the quark spin polarization can be defined
as
\be
\Delta q= q^{+}- q^{-},
\ee
where $q^{\pm}$ can be calculated from the spin
structure of a baryon
\be
\hat B \equiv \langle B|{\cal N}|B \rangle=\langle B|q^+q^-|B \rangle\,. \label{BNB}
\ee
Here $|B\rangle$ is the baryon wave function  and ${\cal N}=q^+q^-$ is the number
operator measuring the sum of the quark  numbers with spin up or down, for example,
\be
q^+q^-=\sum_{q=u,d,s} (n_{q^{+}}q^{+} + n_{q^{-}}q^{-})=n_{u^{+}}u^{+} + n_{u^{-}}u^{-} + n_{d^{+}}d^{+} + n_{d^{-}}d^{-} +
n_{s^{+}}s^{+} + n_{s^{-}}s^{-}\,, \label{number}
\ee
with the coefficients of the $q^{\pm}$ giving the number of
$q^{\pm}$ quarks. The contributions of the quark sea coming from the fluctuation process
in Eq. (\ref{basic}) can be calculated  by substituting for every
constituent quark
\be
q^{\pm} \to \sum P_q q^{\pm} + |\psi(q^{\pm})|^2, \label{sea-q}
\ee
where the transition probability of the emission of a GB from
any of the $q$ quark ($\sum P_q$)  and the transition
probability of the $q^{\pm}$ quark ($|\psi(q^{\pm})|^2$) can be calculated from the Lagrangian. They are expressed as
\[ \sum P_u= a\left( \frac{9+\beta^2+2 \zeta^2}{6} +\alpha^2\right)~~~
{\rm and}~~~
|\psi(u^{\pm})|^2=\frac{a}{6}(3+\beta^2+2 \zeta^2)u^{\mp}+
a d^{\mp}+a \alpha^2 s^{\mp}\,,      \]

\[ \sum P_d= a\left( \frac{9+\beta^2+2 \zeta^2}{6} +\alpha^2\right)~~~
{\rm and}~~~
|\psi(d^{\pm})|^2=a u^{\mp}+
\frac{a}{6}(3+\beta^2+2 \zeta^2)d^{\mp}+ a \alpha^2 s^{\mp}\,,
                                               \]

\[ \sum P_s= a\left( \frac{2 \beta^2+\zeta^2}{3}+2 \alpha^2\right)~~~
{\rm and}~~~
|\psi(s^{\pm})|^2=   a \alpha^2 u^{\mp}+
a \alpha^2 d^{\mp}+\frac{a}{3}(2 \beta^2+\zeta^2)s^{\mp}\,.
 \]

Spin-spin forces,  known to be compatible
\cite{{riska},{chengspin},{prl}} with the $\chi$QM, generate
configuration mixing \cite{{Isgur},{DGG},{yaouanc}} for the
octet baryons which effectively leads to modification
of the spin distribution functions \cite{hd}.
The  general configuration mixing generated by the spin-spin forces
has been discussed in the case of octet baryons \cite{{Isgur},{yaouanc},{full}}.
However, it is adequate
{\cite{{yaouanc},{hd},{mgupta1},{effm2}}} to consider the ``mixed'' octet with mixing only between
$|56,0^+ \rangle_{N=0}$ and the $|70,0^+\rangle_{N=2}$ states, for
example,

\begin{equation}
|B\rangle
\equiv \left|8,{\frac{1}{2}}^+ \right>
= \cos \phi |56,0^+\rangle_{N=0}
+ \sin \phi|70,0^+\rangle_{N=2}\,,  \label{mixed}
\end{equation}
where $\phi$ represents the $|56\rangle-|70\rangle$ mixing and
\bea
 |56,0^+\rangle_{N=0} &=& \frac{1}{\sqrt 2}(\chi^{'} \phi^{'} +
\chi^{''} \phi^{''}) \psi^{s}(0^+)\,, \label{56}   \\
|70,0^+\rangle_{N=2} &=&  \frac{1}{2}[(\phi^{'} \chi^{''} +\phi^{''}\chi^{'})
\psi^{'}(0^+) + (\phi^{'} \chi^{'} -\phi^{''} \chi^{''})\psi^{''}(0^+)]\,.
\label{70}
\eea
In general, the isospin wave functions for the octet baryons ($N$, $\Sigma$, $\Xi$) of the
type $B(xxy)$ are   given as
\be
\phi^{'}_B = \frac{1}{\sqrt 2}(xyx-yxx)\,,~~~
\phi^{''}_B = \frac{1}{\sqrt 6}(2xxy-xyx-yxx)\,,
\ee
whereas for $\Lambda (uds)$  they are  given as
\be
 \phi^{'}_{\Lambda} = \frac{1}{2 \sqrt 3}(usd+sdu-sud-dsu-2uds-2dus)\,,~~~
\phi^{''}_{\Lambda} = \frac{1}{2}(sud+usd-sdu-dsu)\,.
\ee
The spin wave functions are expressed as
\be
\chi^{'} =  \frac{1}{\sqrt 2}(\uparrow \downarrow \uparrow
-\downarrow \uparrow \uparrow)\,,~~~
\chi^{''} =  \frac{1}{\sqrt 6} (2\uparrow \uparrow \downarrow
-\uparrow \downarrow \uparrow -\downarrow \uparrow \uparrow)\,.
\ee
For the definition of the spatial  wave functions ($\psi^{s}, \psi^{'}, \psi^{''})$ as well as the
definitions of the overlap integrals, we  refer the reader to reference {\cite{{yaoubook}}.

The quark polarizations can be calculated from the spin
structure of a given baryon. Using Eqs. (\ref{BNB})
and (\ref{mixed}) of the text, the spin structure of a baryon in the
``mixed'' octet is given as
\be
\hat B \equiv \langle B|{\cal N}|B \rangle={\cos}^2 \phi
{\langle 56,0^+|{\cal N}|56,0^+\rangle}_B +
{\sin}^2 \phi {\langle 70,0^+|{\cal N}|70,0^+ \rangle}_B\,. \label{spin st}
\ee
For the case of $N$, $\Sigma$, $\Xi$ and $\Lambda$, using Eqs. (\ref{56}) and
(\ref{70}), we have
\bea
{\langle 56,0^+|{\cal N}|56,0^+ \rangle}_N&=&\frac{5}{3} u_{+} +\frac{1}{3} u_{-}+
\frac{1}{3} d_{+} +\frac{2}{3} d_{-}\,, \label{56proton} \\
 {\langle 70,0^+|{\cal N}|70,0^+ \rangle}_N&=&\frac{4}{3} u_{+} +\frac{2}{3} u_{-}+
\frac{2}{3} d_{+} +\frac{1}{3} d_{-}\,, \label{70proton}
\eea
\bea
{\langle 56,0^+|{\cal N}|56,0^+ \rangle}_{\Sigma}&=&\frac{5}{3} u_{+} +\frac{1}{3} u_{-}+
\frac{1}{3} s_{+} +\frac{2}{3} s_{-}\,, \label{56sigma} \\
 {\langle 70,0^+|{\cal N}|70,0^+ \rangle}_{\Sigma}&=&\frac{4}{3} u_{+} +\frac{2}{3} u_{-}+
\frac{2}{3} s_{+} +\frac{1}{3} s_{-}\,, \label{70sigma}
\eea
\bea
{\langle 56,0^+|{\cal N}|56,0^+ \rangle}_{\Xi}&=&\frac{5}{3} s_{+} +\frac{1}{3} s_{-}+
\frac{1}{3} u_{+} +\frac{2}{3} u_{-}\,, \label{56xi} \\
 {\langle 70,0^+|{\cal N}|70,0^+ \rangle}_{\Xi}&=&\frac{4}{3} s_{+} +\frac{2}{3} s_{-}+
\frac{2}{3} u_{+} +\frac{1}{3} u_{-}\,, \label{70xi}
\eea
and
\bea
{ \langle 56,0^+|{\cal N}|56,0^+ \rangle}_{\Lambda}&=&\frac{1}{2} u_{+} +\frac{1}{2} u_{-}+
\frac{1}{2} d_{+} +\frac{1}{2} d_{-}+
1 s_{+} +0 s_{-}\,, \label{56lambda} \\
{ \langle 70,0^+|{\cal N}|70,0^+ \rangle}_{\Lambda}&=&\frac{2}{3} u_{+} +\frac{1}{3} u_{-}+
\frac{2}{3} d_{+} +\frac{1}{3} d_{-}+
\frac{2}{3} s_{+} +\frac{1}{3} s_{-}\,, \label{70lambda}
\eea
respectively. Sea contributions can be included by using Eq. (\ref{sea-q}) and the results have been presented in Table \ref{ocsea}. A closer look at the expressions  of
these quantities reveals that the constant factors represent the
Naive Quark Model (NQM) results which do not include the effects of chiral symmetry
breaking. On the other hand, the factors with transition
probability $a$ represent the contribution from the ``quark sea''
in general (with or without SU(3) symmetry breaking).

\begin{table}
\begin{sideways}
\begin{tabular}{cccc}
Baryons & $\Delta u_{{ B}}$ &$\Delta d_{{ B}}$ &
$\Delta s_B$ \\ \hline
$N$ &   $\begin{array}{l} {\cos}^2 \phi \left[\frac{4}{3}-\frac{a}{3} (7+4 \alpha^2+
 \frac{4}{3}\beta^2 +\frac{8}{3} \zeta^2)\right] \\  +{\sin}^2 \phi \left[\frac{2}{3}-\frac{a}{3} (5+2 \alpha^2
+\frac{2}{3}\beta^2 +\frac{4}{3} \zeta^2)\right] \end{array} $
  & $\ba{l} {\cos}^2 \phi \left[-\frac{1}{3}-\frac{a}{3} (2-\alpha^2
-\frac{1}{3}\beta^2 -\frac{2}{3} \zeta^2)\right] \\  +{\sin}^2 \phi \left[\frac{1}{3}-\frac{a}{3} (5+2 \alpha^2
+\frac{2}{3}\beta^2 +\frac{4}{3} \zeta^2)\right]\ea $
 & $-a \alpha^2$ \\[1ex]\\ [-1.5ex]
 $\Sigma$ & $\ba{l}{\cos}^2 \phi \left[\frac{4}{3}-\frac{a}{3}
   (8+3 \alpha^2+ \frac{4}{3}\beta^2+ \frac{8}{3} \zeta^2)\right] \\ +{\sin}^2 \phi \left[\frac{2}{3}-\frac{a}{3} (4+3 \alpha^2+
 \frac{2}{3}\beta^2 +\frac{4}{3} \zeta^2)\right]\ea$ &
$\ba{l}-{\cos}^2 \phi \left[\frac{a}{3}(4-\alpha^2) \right] \\ -{\sin}^2 \phi \left[\frac{a}{3}(2+\alpha^2) \right] \ea$
 & $\ba{l}{\cos}^2 \phi \left[-\frac{1}{3}-\frac{a}{3} (2
\alpha^2-\frac{4}{3}\beta^2 -\frac{2}{3} \zeta^2)\right]\\ +{\sin}^2 \phi  \left[\frac{1}{3}-\frac{a}{3} (4 \alpha^2+
 \frac{4}{3}\beta^2 +\frac{2}{3} \zeta^2)\right]\ea$ \\
  [1ex]\\ [-1.5ex]
 $\Xi$ & $\ba{l}{\cos}^2 \phi \left[-\frac{1}{3}-\frac{a}{3}
   (3 \alpha^2-2- \frac{1}{3}\beta^2 -\frac{2}{3} \zeta^2)\right] \\ +{\sin}^2 \phi \left[\frac{1}{3}-\frac{a}{3} (2+3 \alpha^2+
\frac{1}{3}\beta^2 +\frac{2}{3} \zeta^2)\right]\ea$
&$\ba{l}-{\cos}^2 \phi \left[\frac{a}{3}(4\alpha^2-1) \right]\\ -{\sin}^2 \phi \left[\frac{a}{3}(1+2 \alpha^2) \right] \ea$
&$\ba{l}{\cos}^2 \phi \left[\frac{4}{3}-\frac{a}{3} (7 \alpha^2+
\frac{16}{3}\beta^2 +\frac{8}{3} \zeta^2)\right]\\ +{\sin}^2 \phi \left[\frac{2}{3}-\frac{a}{3} (5 \alpha^2+
\frac{8}{3}\beta^2 +\frac{4}{3} \zeta^2)\right] \ea$ \\
 [1ex]\\ [-1.5ex]
$\Lambda$& $\ba{c} -{\cos}^2 \phi \left[a \alpha^2\right]\\ +{\sin}^2 \phi \left[\frac{1}{3}-\frac{a}{9}(9+6 \alpha^2+\beta^2+
2 \zeta^2)\right]\ea$ &
$\ba{c}-{\cos}^2 \phi \left[a \alpha^2\right] \\+{\sin}^2 \phi \left[\frac{1}{3}-\frac{a}{9}(9+6 \alpha^2+\beta^2+
2 \zeta^2)\right] \ea $ &
$\ba{l}{\cos}^2 \phi \left[1-\frac{a}{3}(6 \alpha^2+
4 \beta^2 +2 \zeta^2)\right]\\+{\sin}^2 \phi \left[\frac{1}{3}-\frac{4}{9}a(3 \alpha^2+2 \beta^2+
\zeta^2)\right] \ea$ \\
\hline
\end{tabular}
\end{sideways}
\caption{The quark spin polarizations for the octet baryons
in the $\chi$CQM.}
\label{ocsea}
\end{table}

\section{Axial Vector Form Factors}

The axial-vector form factors can be expressed in terms of the axial-vector current $A^{\mu,a}$ defined as $\overline{{\bf q}}\gamma^\mu \gamma_5\frac{\lambda^a}{2} {\bf q}$ through the following matrix elements
\be \langle
B(p')|A^{\mu,a}|B(p)\rangle=\bar u(p') \left[
\gamma^\mu \gamma_5 G^i_A(Q^2)+ \frac{q^\mu}{2M_B} \gamma_5 G^i_P(Q^2) \right]  u(p)\,, \label{Amu} \\
\ee where  $M_B$ is the baryon mass,  $u(p)$ ($\bar u(p')$) are the Dirac
spinors of the initial (final) baryon states, respectively. The
four momenta transfer is given as $Q^2 = -q^2$, where $ q \equiv
p- p'$. Here, $\lambda^a$ ($a=1,2,..8$) are the Gell-Mann matrices of SU(3) describing the flavor structure of the 3 light quarks.
It is often convenient to introduce the unit matrix $\lambda^0(=\sqrt{\frac{2}{3}}I)$ in addition to these
matrices. In the present context. we shall need only the matrices having diagonal representation corresponding to the flavor singlet current ($a=0$),  isovector current ($a=3$) and hypercharge axial current ($a=8$) \cite{pcac-ref2}. The functions $G^i_A(Q^2)$ and $G^i_P(Q^2)$ ($i=0,3,8$)
are the axial and induced pseudoscalar form factors respectively. We will ignore the induced pseudoscalar form factors as they not relevant to the present work.

In general, the axial-vector matrix elements have implications for spin structure
\cite{hd,nsweak}. In order to calculate the axial charge as one of the important static property of the form factors at zero momentum transfer, the  singlet and non-singlet combinations of the  spin structure can be related to the weak couplings and can be expressed in terms of the spin polarizations defined in the above section. We have
\bea
g^0_{A,B} &=& \langle B | u^+u^- + d^+d^- + s^+s^-|B \rangle = \Delta u_{B}+\Delta d_{B}+\Delta s_{B} \,, \nonumber \\
g^3_{A,B} &=& \langle B | u^+u^- - d^+d^- |B \rangle = \Delta u_{ B}-\Delta d_{ B} \,, \nonumber \\
g^8_{A,B} &=& \langle B | u^+u^- + d^+d^- + 2 s^+s^-|B \rangle = \Delta u_{ B}+\Delta d_{B}-2\Delta s_{ B} \,. \label{ggg}
\eea
The axial coupling constants $g^3_{A,B}$ and $g^8_{A,B}$ basically correspond to the BSR \cite{bjorken} and the EJSR \cite{ellis}. The axial coupling constant related to the total quark spin content $g^0_{A,B}$ reduces to the EJSR in the $\Delta s=0$ limit.

To compare the $\chi$CQM results with the available experimental data and other model calculations, we can take the case of the quark spin polarizations and the axial coupling constants for the octet baryons at zero momentum transfer. The numerical calculation of the axial-vector coupling constants of the
octet baryons at $Q^2=0$ involves two set of parameters, the  SU(3) symmetry breaking
parameters of $\chi$CQM and the mixing angle $\theta$. The mixing angle $\theta$ is fixed from the
consideration of neutron charge radius \cite{DGG}. The $\chi$CQM
parameters, $a$, $a \alpha^2$, $a \beta^2$, and $a \zeta^2$
represent respectively, the probabilities of fluctuations to
pions, $K$, $\eta$, and $\eta^{'}$. A best fit of $\chi$CQM
parameters can be obtained by carrying out a fine grained analysis
of the spin and flavor distribution functions \cite{hd,hds} wherein as a first step, a gross analysis was carried out to find the limits of the parameters from the well known experimentally measurable quantities while taking into account strong physical considerations. After obtaining the limits, as a second step, a detailed and fine grained analysis was carried out to obtain the best fit. In Table \ref{input}, we summarize the input
parameters and their values.  We would like to mention here that the positive values of $\zeta$ have also been widely used in similar calculations \cite{pcac-ref2}. The sign may not be important for the case of quark spin polarizations in the present context where only $\zeta^2$ is involved but since
this set
of parameters has already been tested for a wide variety of low-energy matrix elements
and is able to give a simultaneous fit to the quantities describing proton spin and
flavour structure including quark flavor distribution functions (anti quark contents, anti up and anti down quark asymmetry, fraction of quark flavors) as well as the magnetic moments of octet and decuplet baryons etc., we use the same set here. A relative negative sign of $\zeta=c_1/c_8$ is required primarily to yield the antiquark $\bar u-\bar d$  asymmetry or the $\bar u/\bar d$
ratio \cite{NMC,baldit,E866} because they involve  $\zeta$.  The results of the quark spin polarizations and the axial coupling constants for the octet baryons at zero momentum transfer using the parameters listed above have been presented in Table \ref{spin-axial}.
\begin{table}
\begin{center}
\begin{tabular}{|c|c|c|c|c|c|}      \hline
Parameter$\rightarrow$ &$\phi$ &$a$ &$\alpha$ &$\beta$ &$\zeta$ \\
\hline
Value & $20^{o}$ & 0.114 & 0.45 & 0.45 & -0.75 \\
 \hline
\end{tabular}
\end{center}
\caption{ Input parameters of the $\chi$CQM used in the analysis.}
\label{input}
\end{table}

\begin{table}
\begin{center}
\begin{tabular}{|c|c|c|c|c|}
\hline  Quantity  & $N$ & $\Sigma$  & $\Xi$ & $\Lambda$ \\ \hline
$\Delta u_B$ & 0.904  & 0.881 &$-0.329$ & 0.002 \\
$\Delta d_B $ & $-0.362$  & $-0.137$ & 0.00 & 0.002 \\
$\Delta s_B$ & $-0.023$ & $-0.252$ & 1.109 & 0.805 \\
$g^0_{A,B}$ & 0.519& 0.492  & 0.780 & 0.809 \\
$g^3_{A,B}$ & 1.266& 1.018  & $-0.329$ & 0.00 \\
$g^8_{A,B}$ & 0.588 & 1.248  & $-2.547$ & $-1.606$ \\\hline
\end{tabular}
\caption{ The $\chi$CQM results for the quark spin polarizations and the axial coupling constants for the $N$, $\Sigma$, $\Xi$ and $\Lambda$ octet baryons.}\label{spin-axial}
\end{center}
\end{table}

The present experimental situation \cite{PDG}, in terms
of the quark spin polarizations, $\Delta u$, $\Delta d$ and $\Delta s$ for the case of $N$,
is summarized as follows:
\bea
\Delta u^{\text {expt}}_N& =& 0.85\pm 0.05,~~~~
\Delta d^{\text {expt}}_N = -0.41 \pm 0.05,~~~~
\Delta s^{\text {expt}}_N = -0.07 \pm 0.05 \,,  \nonumber \\
g^{0~\text{expt}}_{A,N} &=&0.30\pm 0.06,~~~~
g^{3~\text{expt}}_{A,N}= 1.267 \pm 0.0025,~~~~
g^{8~\text{ expt}}_{A,N}= 0.588 \pm 0.033 \,,
\eea
The NQM, which is
quite successful in explaining a good deal of low energy data
\cite{Isgur,DGG,yaouanc},
has the following predictions for the above mentioned quantities
\bea
\Delta u_N &=& 1.33, ~~~\Delta d_N = -0.33, ~~~ \Delta s_N = 0\,,  \nonumber \\
g^0_{A,N} &=& 1, ~~~g^3_{A,N}= 1.66,  ~~~g^8_{A,N}= 1\,.
\eea
The disagreement between the
NQM predictions and the DIS measurements was broadly characterized as ``proton spin crisis''.
The results of $\chi$CQM for the case of $\Delta u_N$, $\Delta d_N$, $\Delta s_N$, $g^3_{A,N}$ and $g^8_{A,N}$ are more or less in agreement with data. This not only justifies the success of $\chi$CQM but also strengthens our conclusion regarding  the qualitative and quantitative role of the ``quark sea'' in right direction. For the case of $g^0_{A,N}$, the NQM results show that the valence quarks of the nucleon carry only about 1/3 of the nucleon spin as obtained in the experiment. The $\chi$CQM result comes out to be 0.519 which is better than the results of NQM but still shows a large deviation from data. A detailed understanding of the deep inelastic results as well as the dynamics of the constituents of the nucleon
constitute a major challenge for any model trying to explain the nonperturbative regime of QCD. In this context, it has been shown recently in a chiral constituent quark potential model that it is possible to describe the singlet axial nucleon coupling if consistent axial exchange currents are taken into account \cite{pcac-ref2,buchmann-spin,thomas-spin}. Because of angular momentum conservation, this reduction of the quark spin is compensated
by orbital angular momentum carried by the same nonvalence quark degrees of freedom.

The $Q^2$ dependence of the axial-vector form factors have been experimentally investigated from the quasi elastic neutrino scattering \cite{antineutrino1,antineutrino2} and from the pion electroproduction \cite{pion-electro}. The dipole form of parametrization has been conventionally used to analyse the axial-vector form factors
\be
G^i_{A,B}(Q^2)=\frac{g^i_{A,B}(0)}{\left( 1+\frac{Q^2}{M_{A}^2}\right)^2},
\label{dipole}\ee
where $g_A^0(0)$, $g_A^3(0)$ and $g_A^8(0)$ are the isovector axial-vector coupling constants at zero momentum transfer. For the axial mass $M_A$, a global average as extracted from neutrino scattering
experiments is $M_A = (1.026 \pm 0.021)$GeV \cite{neutrino}. Another recent analysis
finds a slightly smaller value
$M_A = (1.001 \pm 0.020)$GeV \cite{MA}. However, in the present work
we have used the most recent value obtained by the MiniBooNE Collaboration $M_A = 1.10^{+0.13}_{-0.15}$GeV \cite{miniboone}. The axial mass can be taken as free parameter and adjusted to experiment \cite{buchmann-axial-mass}. Since experimental data is available only for the nucleon axial coupling constants, we have used the same value of the axial mass for all the octet baryons. The axial masses corresponding to $\Sigma$, $\Xi$ and $\Lambda$ are expected to be larger than that of the nucleon which will in turn lead to slightly larger values of the axial-vector form factors in magnitude. The overall behavior of the form factors however will not be affected by this change.

\begin{figure}
\includegraphics {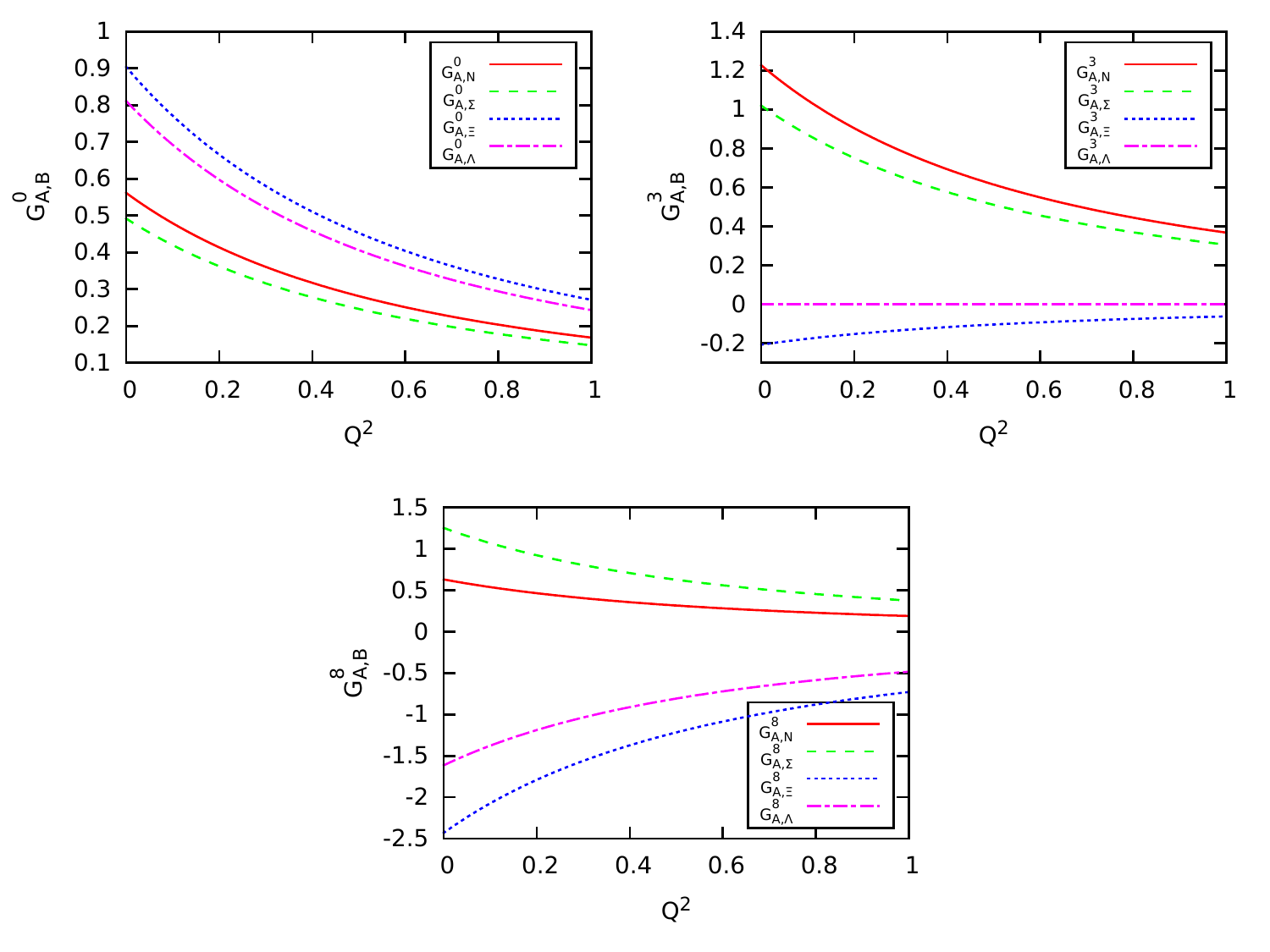}
\caption{(color online). Form factors  for the baryons $N,\,\Sigma,\,\Xi,\,\Lambda$ plotted as function of  $Q^2$.}
\label{fig1-g038}
\end{figure}

After having incorporated $Q^2$ dependence in the axial-vector form factors, we now discuss the variation of all the $Q^2$ dependent
quantities in the range $0\leq Q^2\leq 1$. In Fig. \ref{fig1-g038}, we have presented the singlet and non-singlet axial-vector form factors of the octet baryons $N$, $\Sigma$, $\Xi$ and $\Lambda$. From a cursory look at the plots, one can easily describe some
general aspects of the sensitivity to $Q^2$ for the form factors. The
sensitivity of the singlet and non-singlet form factors for different baryons varies as
\bea G^0_{A,\Xi} > G^0_{A,\Lambda} >  G^0_{A,N} > G^0_{A,\Sigma}, \nonumber \\
G^3_{A,N} > G^3_{A,\Sigma} >  G^3_{A,\Xi} > G^3_{A,\Lambda}, \nonumber \\
G^8_{A,\Xi} > G^8_{A,\Lambda} >  G^8_{A,N} > G^8_{A,\Sigma}.
\eea
The behaviour of the form factors for $\Xi$ and $\Lambda$ is similar to each other. This may possibly due to the presence of more strange quarks in the valence structure. On the other hand,  the form factors for $N$ and $\Sigma$, which have the dominance of $u$ quarks in the valence structure, show similar variation with $Q^2$. This can be easily seen from Fig. \ref{fig1-g038} and this is true for $G^0_{A,B}$, $G^3_{A,B}$ as well as $G^8_{A,B}$. Another  important observation for the case of $G^0_{A,B}$ form factors is that it falls off rapidly with the increase of $Q^2$ for all the octet baryons $N$, $\Sigma$, $\Xi$ and $\Lambda$. However, for the case of  $G^3_{A,B}$ and $G^8_{A,B}$, the $N$ and $\Sigma$ form factors fall off with increasing $Q^2$ whereas the $\Xi$ and $\Lambda$ form factors increase with increasing $Q^2$. The case of $G^3_{A,\Lambda}$ is particularly interesting because of its flavor structure which has equal numbers of $u$, $d$, and $s$ quarks in its valence structure. Unlike the other octet baryons, where the form factors decrease or increase continuously
with the $Q^2$ values, the form factor in this case has no $Q^2$ dependence.

\begin{figure}
\includegraphics {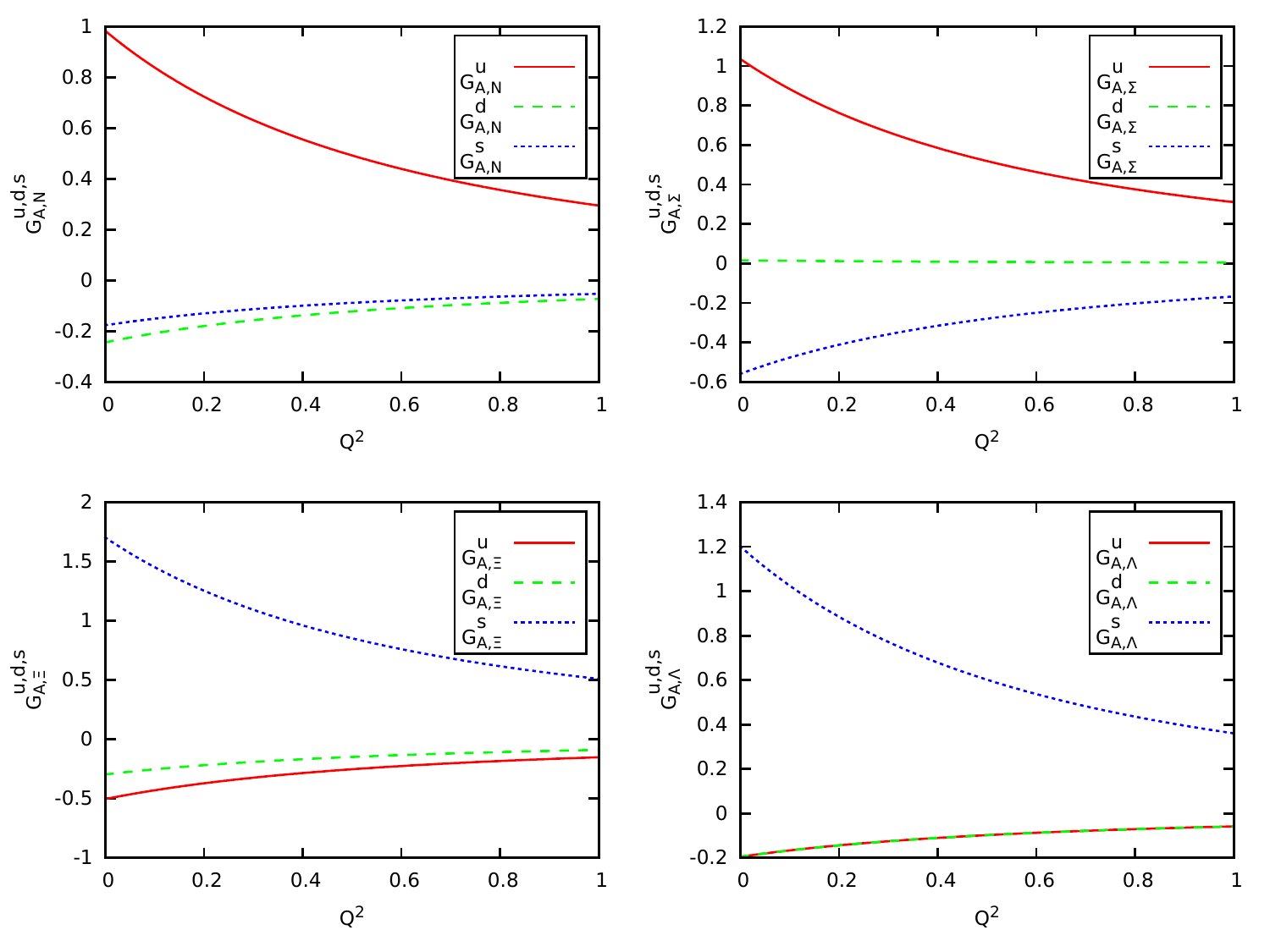}
\caption{(color online). The explicit flavor form factors for the baryons $N,\,\Sigma,\,\Xi,\,\Lambda$ plotted as function of  $Q^2$.}
\label{fig2-uds}
\end{figure}

Since the constituent quarks are spatially extended particles \cite{buchmann-axial-mass,q2range}, they themselves have axial form factors. The role of non-valence quarks in the spin structure can be studied in detail by calculating the flavor
axial-vector form factors using the dipole form of parametrization (Eq. (\ref{dipole}). These can be expressed in terms of the  singlet and non-singlet combinations of the  spin
structure as follows
\bea
G^u_{A,B} &=& \frac{1}{3} G^0_{A,B}+\frac{1}{2} G^3_{A,B}+\frac{1}{2\sqrt{3}} G^8_{A,B}\,, \nonumber \\
G^d_{A,B} &=& \frac{1}{3} G^0_{A,B}-\frac{1}{2} G^3_{A,B}+\frac{1}{2\sqrt{3}} G^8_{A,B} \,, \nonumber \\
G^s_{A,B} &=& \frac{1}{3} G^0_{A,B}-\frac{1}{\sqrt{3}} G^8_{A,B} \,. \label{uds-formfactors}
\eea
In Fig. \ref{fig2-uds}, we have plotted the explicit $u$, $d$, and $s$ quark flavor
contributions for each of the octet baryon axial-vector form factors. The plots clearly project out the valence quark structure of the baryon. For example, since $N$ is dominated by $u$ quark it is clear from the plot of $G^{u,d,s}_{A,N}$ that the  $G^{u}_{A,N}$ dominates and $G^{d}_{A,N}$, $G^{s}_{A,N}$ has a comparatively smaller contribution.  The important observation in this case is the non-zero contribution of the $s$ quarks. Even though there are no $s$ quarks in the valence structure the contribution of $G^{s}_{A,N}$ implies a presence of ``quark sea'' which is even more at zero momentum transfer. It is also evident from the figure that the valence quark distribution is spread over the entire $Q^2$ region and as the value of $Q^2$ increases, the sea contributions decrease and at even higher values of $Q^2$ (not presented here), the contributions should be completely dominated by the valence quarks.
Further, for the case of $G^{u,d,s}_{A,\Sigma}$ and $G^{u,d,s}_{A,\Xi}$, where the valence structure is dominated by the $u$ and $s$  quarks, we find a significant contribution from them. In these form factors, the small but significant $G^{d}_{A}$ can have important implications for the
role of sea quarks at low $Q^2$. Finally, the $G^{u,d,s}_{A,\Lambda}$, even after having equal contributions from the $u$, $d$, and $s$ quarks, does not show a symmetric behaviour. The $G^{s}_{A,\Lambda}$ clearly dominates over  $G^{u}_{A,\Lambda}$ and $G^{d}_{A,\Lambda}$ which is expected because of the $u$ and $d$ quarks also contribute towards $G^{u,d,s}_{A,\Lambda}$ through quark fluctuations. It is interesting to note that the valence and sea quark distributions
contribute in the right direction to give an excellent overall fit to the axial-vector form factors where experimental data is available. This can perhaps be substantiated further by a measurements for the other octet baryons.

\begin{figure}
\includegraphics {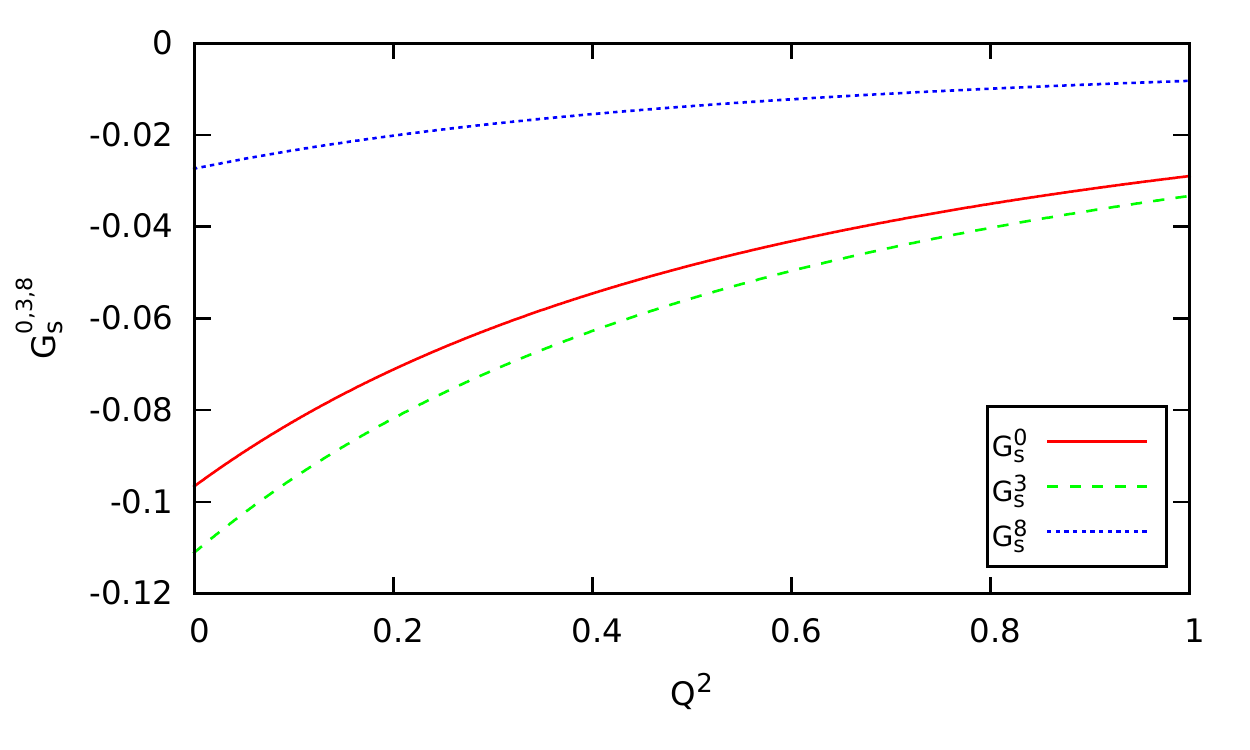}
\caption{(color online). The strange form factors for the nucleon plotted as function of  $Q^2$.}
\label{fig3-gsn}
\end{figure}

It is well known that, for the case of nucleon, the strange quarks contribute to the spin
polarizations of $u$ and $d$ quarks apart from contributing to the strange spin
polarization. This is because of the presence of the non-valence ``quark sea'' (Eq. (\ref{basic})). In this context, the axial-vector matrix elements will have implications for the strangeness contribution to the nucleon as well as for the effects of chiral symmetry
breaking. We can calculate $G_s^0(Q^2)$, $G_s^3(Q^2)$ and $G_s^8(Q^2)$ for the case of  $N$ from Eq. (\ref{ggg}) and Table \ref{ocsea} by dropping the constant factors. The factors with $a\alpha^2$, $a\beta^2$ and $a\zeta^2$ include the effects of chiral symmetry breaking
as well as SU(3) symmetry breaking and give the contribution coming from the ``quark sea''. In particular, they give the contribution of strange quarks to the nucleon spin. The explicit strangeness contribution for the other octet baryons is not so significant because of the presence of strange quarks in their valence structure.
In Fig. \ref{fig3-gsn}, we have presented the results for $G_s^0(Q^2)$, $G_s^3(Q^2)$ and $G_s^8(Q^2)$ for the case of $N$. We find that the magnitude of $G_s^0(Q^2)$ and $G_s^8(Q^2)$ fall off with the increasing value of $Q^2$ whereas  $G_s^3(Q^2)$ has a weak  $Q^2$  dependence. For the sake of completeness we have also presented the numerical values of the explicit strangeness contribution to $\Delta u$, $\Delta d$ and the axial coupling constants at $Q^2=0$ for the case of $N$ in Table \ref{gsn0}. The contribution of $\Delta s$ is coming purely from the quark sea and has already been presented in Table \ref{spin-axial}. It is clear from the table that there is a significant contribution of non-valence quarks in $\Delta u_s$, $g^0_s$ and $g^3_s$.  These quantities not only provide a direct method to determine the presence of a significant amount of quark sea  but also impose important constraint on a model that attempts to describe the origin of the quark sea. A small but significant contribution of strangeness in the nucleon has already been indicated by SAMPLE at MIT-Bates \cite{sample}, G0 at JLab \cite{g0},
PVA4 at MAMI \cite{a4} and HAPPEX  at JLab \cite{happex}. A determination of $G_A^s$ at low values of $Q^2$ \cite{miniboone} would permit a determination of strange spin polarization $\Delta s$ which is otherwise zero in the case of nucleon. The  strange quarks contribute through the quark sea generated by the
chiral fluctuations and any refinement in the case of the strangeness dependent quantities would have important implications for the basic tenets of $\chi$CQM.

\begin{table}
\begin{center}
\begin{tabular}{|c|c|c|c|c|c|}
\hline  Quantity $\rightarrow$  & $\Delta u_s$ & $\Delta d_s$ & $g^0_s$ & $g^3_s$ & $g^8_s$ \\ \hline
 NQM & 0 & 0 & 0 & 0 & 0  \\
$\chi$CQM & $-0.092$ & $0.013$ & $-0.102$ & $-0.105$ & $-0.033$  \\ \hline
\end{tabular}
\caption{ The NQM and $\chi$CQM results for the explicit strangeness contribution to spin polarizations and the axial coupling constants at $Q^2=0$ for the case of $N$.}\label{gsn0}
\end{center}
\end{table}

To summarize, the  chiral constituent quark model ($\chi$CQM) is able to phenomenologically estimate the quantities having implications for chiral symmetry
breaking and SU(3) symmetry breaking. In particular, it provides a fairly good description of the axial-vector form factors of the low lying octet baryons ($N$, $\Sigma$, $\Xi$ and $\Lambda$), for example, the singlet ($g^0_{A}$) and non-singlet ($g^3_{A}$ and $g^8_{A}$) axial-vector coupling constants expressed as combinations of the spin polarizations at zero momentum transfer. In order to enlarge the scope of $\chi$CQM, we have used the conventional dipole form of parametrization to analyse the $Q^2$ dependence of the axial-vector form factors ($G^0_{A}(Q^2)$, $G^3_{A}(Q^2)$ and $G^8_{A}(Q^2)$). To understand the role of chiral symmetry breaking and the significance of non-valence quarks in the nucleon structure, the implications of hidden strangeness component have been studied for the strange singlet and non-singlet contents ($G_s^0(Q^2)$, $G_s^3(Q^2)$ and $G_s^8(Q^2)$) of the nucleon. The $\chi$CQM is able to give a qualitative and quantitative description of the axial-vector form factors. The significant contribution of the strangeness is also consistent with the recent available experimental results.

In conclusion, we would like to state that chiral symmetry breaking and SU(3) symmetry breaking play an important role in understanding the spin structure of the baryon and is the key to describe the hidden strangeness content of the nucleon in the nonperturbative regime of QCD where the  constituent quarks and the weakly interacting Goldstone bosons constitute the appropriate degrees of freedom at the leading order. The future experiments to measure the axial-vector form factors will not only provide a direct method to determine the presence of appropriate amount of quark sea  but also impose important constraint on the parity-violating asymmetries in different kinematical regions. Several groups, for example, Miner$\nu$a are contemplating the possibility of performing the high precision measurements over a wide $Q^2$ region in the near future.

\section*{ACKNOWLEDGMENTS}

HD would like to thank Department of Science and Technology, Government of India for financial support.

\end{document}